\documentclass[reprint,superscriptaddress,amsmath,amssymb,aps,prl]{revtex4-2}
\usepackage{graphicx}
\usepackage{mathtools}
\usepackage{xcolor}
\usepackage{physics}
\usepackage{amsmath}
\usepackage{float}
\usepackage{comment}
\usepackage{hyperref}
\usepackage[caption=false]{subfig}
\usepackage{lipsum}
\usepackage{bbold}
\usepackage{feynmp}
\usepackage{natbib}
\usepackage{dcolumn}
\usepackage{bm}

\newcommand{\blue}[1]{#1}

\begin{document}

\title{\blue{Quantum-to-classical crossover in the spin glass dynamics of cavity QED simulators}}

\author{Hossein Hosseinabadi}
\email[hhossein@uni-mainz.de]{}
\affiliation{Institut f{\"u}r Physik, Johannes Gutenberg-Universit{\"a}t Mainz, 55099 Mainz, Germany}
\author{Darrick E. Chang}
\affiliation{ICFO—Institut de Ci{\`e}ncies Fot{\`o}niques, The Barcelona Institute of Science and Technology, 08860 Castelldefels, Spain}
\affiliation{ICREA—Instituci{\'o} Catalana de Recerca i Estudis Avan{\c c}ats, 08015 Barcelona, Spain}
\author{Jamir Marino}
\affiliation{Institut f{\"u}r Physik, Johannes Gutenberg-Universit{\"a}t Mainz, 55099 Mainz, Germany}

\begin{abstract}
\blue{By solving the quench dynamics of a frustrated many-body spin-boson problem, we investigate the role of spin size on the dynamical formation of spin glass order. In particular, we observe that quantum and classical spin glasses exhibit markedly different evolution. The former displays a quick relaxation of magnetization together with an exponential dependence of the spin glass order parameter on spin size, while the latter has long-lasting prethermal magnetization and a spin glass order parameter independent of spin size. The quantum-to-classical crossover is sharp and occurs for relatively small spins, highlighting the fragility of the quantum regime. Furthermore, we show that spin glass order is resonantly enhanced when the frequency of the bosonic mediators of the interactions approaches the value of the transverse field. Our predictions are relevant for all spin glass systems with $SU(2)$ degrees of freedom away from equilibrium, and can be examined in recently developed multi-mode cavity QED experiments.}
\end{abstract}


\maketitle

\emph{Introduction} -- Spin glasses (SG) are frozen states of spins due to competing interactions generated by strong static disorder in their host materials \cite{SG_RMP,mezard1987spin}. Disorder prohibits the formation of long-range ferromagnetic (FM) or anti-ferromagnetic (AFM) orders and at the same time, hinders the melting of the frozen state by fluctuations and the formation of a paramagnetic (PM) state. SGs occupy a distinctive position in the field of disordered systems for their significance in our understanding of ergodicity breaking and for their broad range of applications from neural networks to optimization problems encompassing dynamics of complex biological systems and quantum information~\cite{stein2013spin}. 
While disordered magnetic materials have historically been the main context for investigating SG, state-of-the-art quantum simulators possess the capacity to fabricate  strongly correlated systems under  controlled settings~\cite{PRXQuantum.2.017003, gross2017quantum}. This presents a unique opportunity to understand the interplay of disorder, fluctuations and interactions in the emergence of complex phases of matter, free from the complications and microscopic details of real-world samples.
\begin{figure}[!t]
    \centering
    \includegraphics[width=.96\linewidth]{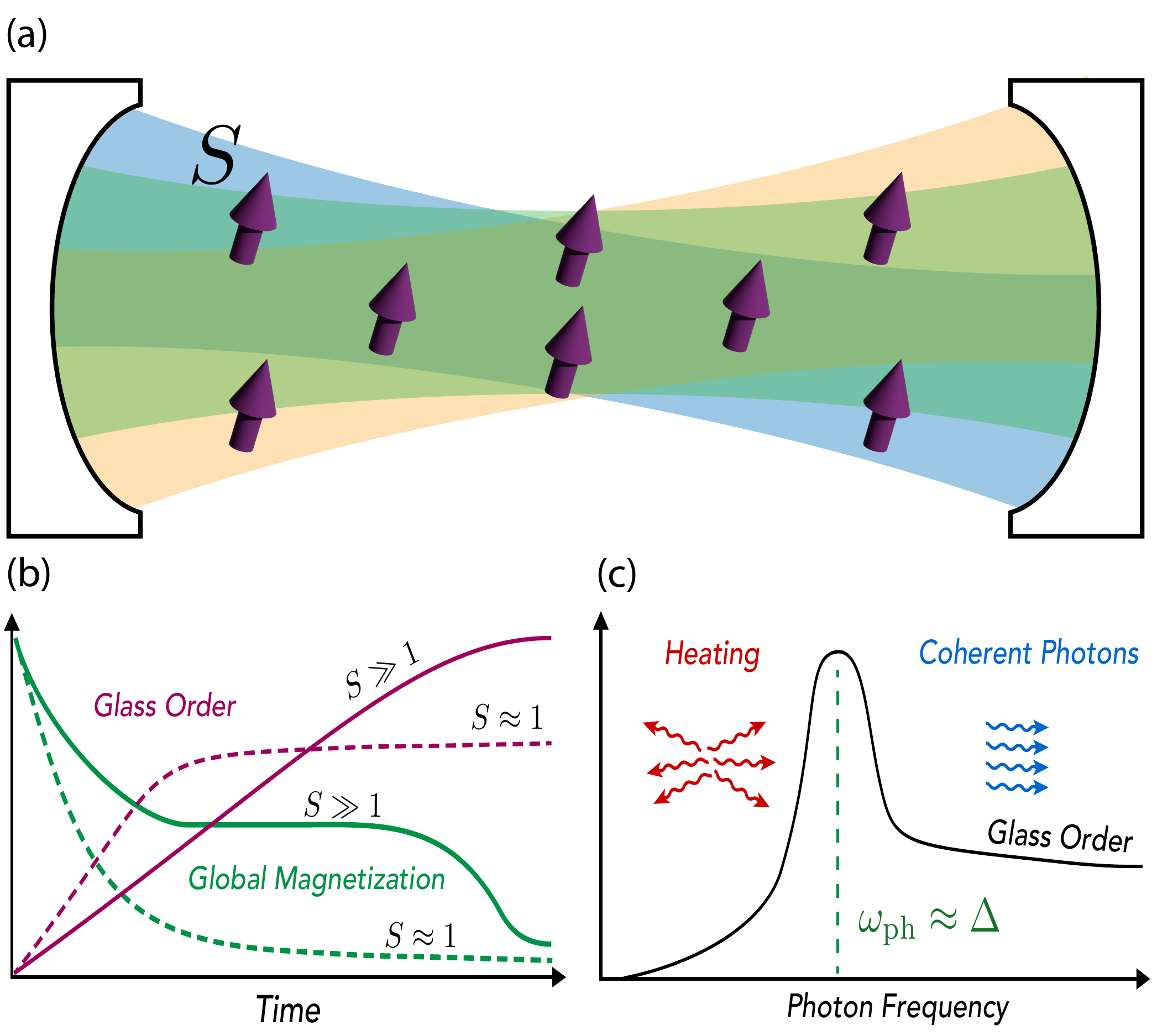}
    \caption{\blue{(a) Schematics of the system considered in this work. Clusters of two-level atoms can be modeled as spins of size $S$. Coupling to a multiple cavity modes of different spatial profile results in frustrated long-range interactions among large spins, which can stabilize a spin glass phase. (b) Generic behavior of the system after an interaction quench. SG order parameter grows much faster for small spins compared to large spins, but to a smaller value, due to stronger quantum fluctuations. (c) Resonantly enhanced SG order, as the frequency of interaction mediator (photons) approaches the atomic splitting $\Delta$.}} 
    \label{fig:cartoon}
\end{figure}

\blue{A particularly interesting objective is to discern the boundary separating quantum and classical SG. Numerous quantum many-body systems admit effective classical descriptions ~\cite{Golubev_decoherence98,altland2010condensed} even though their constituents have quantum behavior. This quantum-to-classical (QC) crossover is usually a result of the competition between decoherence, enhanced by the system-environment coupling or the temperature\cite{Arndt_RydbergQC91,Zurek_decoherence2003}, and quantum fluctuations, which are stronger at lower dimensions and higher densities~\cite{leggett2006,Ra_qc2013}. In spin systems, the size of spins ($S$) is another control parameter for the strength of quantum fluctuations~\cite{chubukov1991quantum,altland2010condensed,Coletta_largeS2016,yamamoto_qc2021}. From an experimental point of view, adjusting $S$ is often a formidable task, primarily because it is intrinsic to the material being investigated.} The cavity QED experiment reported in~\cite{vaidya2018tunable,PRXQuantum.4.020326,kollar2015adjustable,PhysRevLett.122.193601} provides an instance of a platform capable of quantum-simulating a SG (see in particular the note at the end of this Letter), with a high degree of control over parameters of the system, including the size of spins. The experiment consists of an ensemble of ultracold atomic clusters trapped by optical tweezers,  coupled to each other via long-range frustrated interactions mediated by cavity photons (Fig.~\ref{fig:cartoon}a). This platform has been shown to be a simulator of associative memory and SG phases~\cite{Hopfield,Amit_PRL85,Gopal_PRL11,Strack_PRL11,Buchhold_PRA13,Rotondo_Hopfield2015,Fiorelli_PRL20,Marsh_PRX21,Marsh2023}. However, accessing the broad spectrum of dynamical responses in SG systems, and in particular this platform, would require to solve a frustrated dissipative quantum many-body system with retarded (photon-mediated) interactions. In general, there have been limited attempts thus far to solve for the time evolution of text-book models of SG and similar disordered systems~\cite{cugliandolo1999real,kennett2001aging,biroli2002out,PhysRevLett.125.260405,thomson2020quantum,pappalardi2020quantum,Bera_QuantClassSG2022,Clayes_PRA24}.
~\\

\blue{\emph{Outline of results} -- In this Letter we make a significant stride  in addressing far-from-equilibrium dynamics in quantum SGs and in particular, those realized by frustrated light-matter interactions in cavity QED, by providing a non-perturbative solution of their  long-time dynamics. We show that while SG order prevails for all spin sizes, it displays qualitatively different dynamical responses for different values of $S$ (Fig.~\ref{fig:cartoon}b). For small $S$, which we label as quantum SG, we observe weak aging where the memory of the initial state is drastically blurred by quantum fluctuations, alongside with a quick growth of the SG order parameter to a finite but small value. For large $S$, the system displays stronger signatures of aging akin to classical spin glasses, with a slow growth of the SG order parameter. The QC crossover is quite sharp and the system fully classicalizes at spin sizes only few times larger than $\hbar$, the fundamental unit of angular momentum, with an exponential dependence of SG order parameter on $S$, demonstrating the fragility of quantum SG against decoherence. $S$ can be controlled in cavity QED experiments by adjusting the atomic load of each cluster trapped by optical tweezers and, in principle, can be tuned down to the quantum limit $S=1/2$ by activating Rydberg blockade within each cluster~\cite{Marsh2023}. Moreover, we show that tuning the photon gap to resonance with the atomic splitting enhances SG order (Fig.~\ref{fig:cartoon}c). We note that this is a non-trivial result due to the trade-off between atom-atom interactions and incoherent generation of photons, both of which are enhanced close to the resonance~\cite{Babadi_SC2017,Kelly2021,eckhardt_resonance2024}. While the former stabilizes SG, the latter leads to heating, which generally has a detrimental effect on ordered phases of matter. We show that below the resonance, heating dominates and SG order quickly approaches zero.}\\

\emph{ Model} -- Inspired by cavity QED experiments~\cite{vaidya2018tunable,PRXQuantum.4.020326,kollar2015adjustable,PhysRevLett.122.193601}, we consider a system of $N$ clusters, each one containing $N_s$ two-level atoms, placed inside a multi-mode cavity with $M$ nearly degenerate modes.  Atoms are encoded by the Pauli operators $\sigma_{i\lambda}$, with cluster $1\leq i \leq N$ and atom indices $1\leq \lambda \leq N_s$.   Evolution of the system is given by a quantum master equation $\partial_t \rho = -i \comm{H}{\rho} + \sum_{\alpha=1}^M \mathcal{D}\qty[a_\alpha]\rho$ for a dissipative spin-boson model~\cite{Strack_PRL11,Buchhold_PRA13} where

\begin{figure}[!t]
    \centering
    \subfloat[\label{fig:s_vs_S}]{\includegraphics[width=.498\linewidth,trim={0 0.3cm 0 0.0cm},clip]{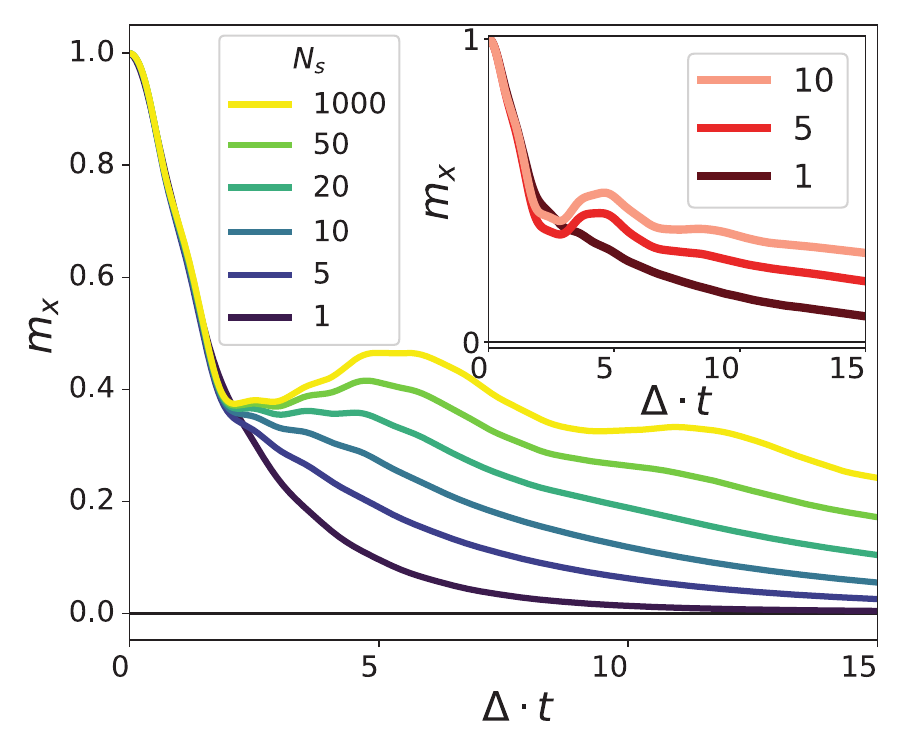}}
    \subfloat[\label{fig:C_vs_g}]{\includegraphics[width=.498\linewidth,trim={0 0.3cm 0 0.0cm},clip]{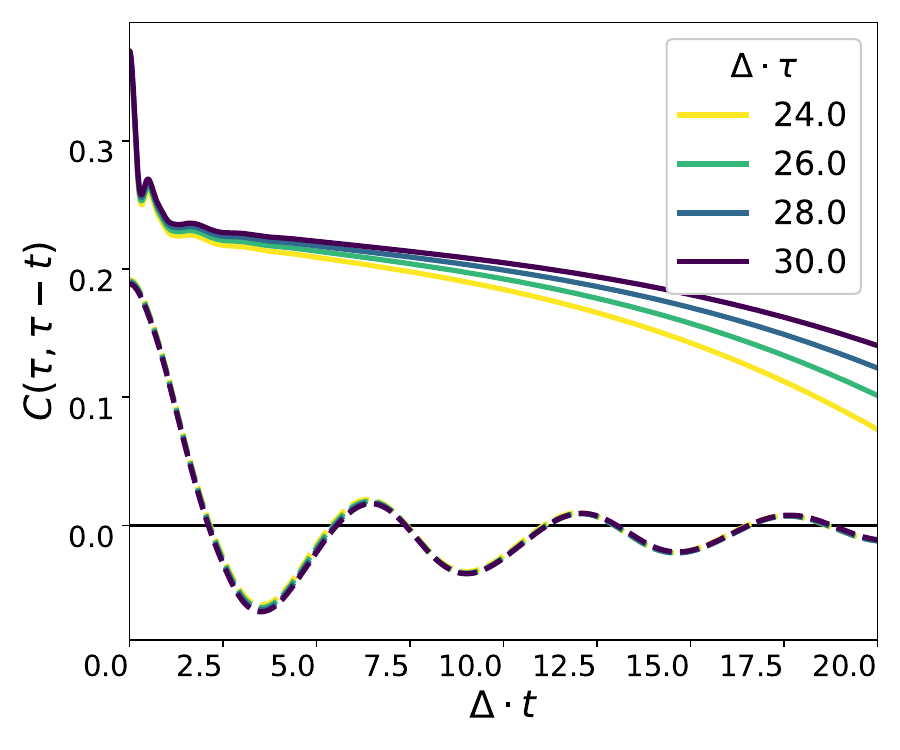}}
    \caption{\blue{(a) Dependence of average magnetization $m_x$ on spin size ($N_s$) for quenches into SG phase with $g/\Delta\approx 2.0$. Inset: results of semi-classical (DTWA) calculations. (b) Spin correlation function at different waiting times after the quench for $N_s=5$. For quenches to PM phase ($g/\Delta\approx 0.18$), correlations decay quickly with $t$  and show weak sensitivity to the waiting time $\tau$ (dashed lines). For quenches to SG phase ($g/\Delta\approx0.4$), correlations remain finite at long time separations and the system retains the memory of its past (solid lines).}} 
    
\end{figure}

\begin{equation}\label{H}
    H=\frac{\Delta}{2} \sum_{i,\lambda} \sigma^z_{i\lambda}+ \omega \sum_{\alpha} a^\dagger_\alpha a_\alpha  + \sum_{\substack{i,\lambda,\alpha}} g_{\alpha i} \qty(a_\alpha+a_\alpha^\dagger) \sigma^x_{i\lambda},
\end{equation}
and $\mathcal{D}\qty[a_\alpha]\rho=\kappa \qty(2 a_\alpha \rho a_\alpha^\dagger - \acomm{a_\alpha^\dagger a_\alpha}{\rho})$. The couplings between the atoms and the photon modes of the cavity are spatially dependent and uncorrelated  from each other, which justifies their modelling via  random couplings~\cite{Strack_PRL11,Buchhold_PRA13,wierzchucka2023integrability}. Accordingly, $g_{\alpha i}$ are assumed to be random and chosen from a Gaussian distribution with $\overline{g_{\alpha i}}=0$ and $\overline{g_{\alpha i}g_{\beta j}}=\delta_{\alpha\beta}\delta_{ij}g^2/(N+M)N_s$. Couplings for spins in the same cluster are similar as we assume that the spatial size of each cluster is smaller than the wavelength of cavity modes. Starting from the same initial state for all spins, each cluster is equivalent to a single spin $2\hat{S}_i \equiv \sum_{\lambda}\hat{\sigma}_{i\lambda}$ with amplitude $S=N_s/2$.
The parameter $S$ can be tuned by loading   few or many atoms in each cluster, and it   dictates the strength of quantum fluctuations. For instance, at large $S$ each cluster would be effectively described by a classical angular momentum, since its quantum noise would scale down as    $1/S$~\cite{kirton2019introduction, Fiorelli_PRL20, Marsh2023}. Integrating out photons leads to frustrated spin interactions~\cite{damanet2019atom,palacino2021atom,jager2022lindblad}, and Eq.~\eqref{H} is mapped to the Hopfield model (HM) \cite{Hopfield}. The HM has a PM ground-state for sufficiently small $g$. For $g$ larger than a critical value $g_c$, a phase transition occurs and the ground-state crucially depends on the ratio $\eta\equiv M/N$ \cite{Amit_PRL85}. For $\eta < \eta_c\sim O(10^{-1})$, the system is in the memory retrieval phase \cite{Gopal_PRL11,Fiorelli_PRL20,Marsh_PRX21}, which is a FM in disguise~\cite{SG_RMP} with multiple superradiant/FM ground-states. For $\eta>\eta_c$ \cite{Amit_PRL85}, frustration dominates and turns the system into a spin glass \cite{Strack_PRL11,Buchhold_PRA13,Rotondo_Glass2015,Marsh2023} described by the Sherrington-Kirkpatrick model \cite{SK_PRL75,Ray_PRB89}. For a correct description of quench dynamics or in the limit of dynamically active photons $\omega \approx \Delta$, photons have to be retained as dynamical degrees of freedom and cannot be integrated out~\cite{damanet2019atom, palacino2021atom,jager2022lindblad}. For this, we resort to non-equilibrium quantum field theory (NEQFT), covered comprehensively in an accompanying article~\cite{HosseinLong}.\\ 

\begin{figure}[!t]
    \centering
    \hspace{-20pt}\includegraphics[width=.8\linewidth,trim={0 0.3cm 0 0.0cm},clip]{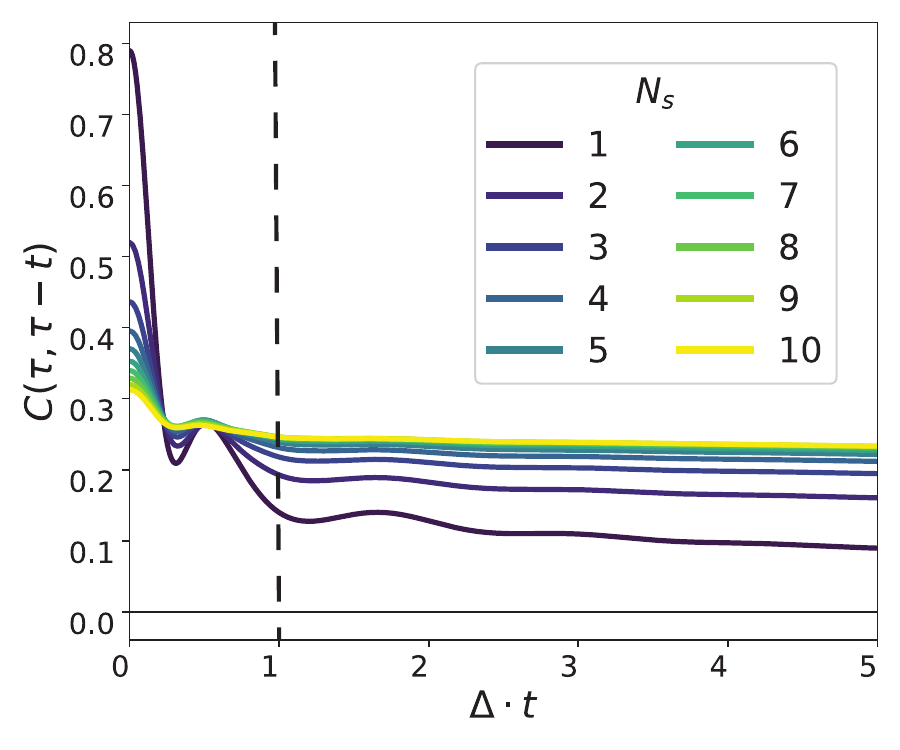}
    \caption{
    \blue{Dependence of temporal correlations on spin size in SG phase. Short time correlations are stronger for small spins, due to the smaller size of local Hilbert space (cf. main text). Small spins are susceptible to quantum fluctuations and their correlations are sharply suppressed for $\Delta t\gtrsim 1$ (marked by dashed line), leading to a quantum SG with weak aging. Large spins are robust against quantum fluctuations and realize a classical SG with stronger aging. The waiting time and the coupling are $\Delta\tau=30$ and $g/\Delta\approx0.4$. }}
    \label{fig:C_vs_S}
\end{figure}

\emph{Dynamics of SG formation} -- Throughout this Letter we take $N=M$, ensuring the existence of SG phase for sufficiently strong couplings \cite{Amit_PRL85,Strack_PRL11}. For now, we take $\omega_c/\Delta=5.0$ and also introduce a non-vanishing cavity loss $\kappa/\Delta=0.5$, consistent with realistic systems.
The initial state is the vacuum state for photons and a uniform product state for spins specified by unit vector $(\sin{\theta_0},\cos{\theta_0})$ in the $xz$-plane. We let the system evolve after suddenly switching on the coupling at $t=0$. For sufficiently weak couplings, the system is a PM and spins precess around the $z$-axis and dephase due to disorder, similar to an under-damped oscillator. As the coupling is increased, the system undergoes a phase transition  into a SG. The simplest manifestation of the transition is the over-damped relaxation of global magnetization to zero: $m_x\equiv \overline{\expval{\sigma^x_i}}$ is shown in Fig.~\ref{fig:s_vs_S}, displaying a crucial dependence on spin size. At early times, spin dynamics are insensitive to $S$, and $m_x$ quickly collapses to a finite value. After this point, trajectories for different spin sizes start to depart, and relaxation becomes slow for larger spins. This behavior can be attributed to the inefficiency of quantum fluctuations to erase the memory of the initial state for larger spins where a few applications of the transverse field ($\Delta S^z$) only changes the amplitude of $m_x$, in contrast to small spins which can be entirely flipped over similar timescales. The long-lasting, transient magnetic order for large $N_s$ is reminiscent of prethermalization, where a system approaches true equilibrium over long times due to disorder or an extensive number of nearly-conserved local quantities \cite{Berges_PRL04,Bertini_prethermal2015,Babadi_PRX15,marino2022dynamical}. \blue{In the inset of Fig.~\ref{fig:s_vs_S} we have shown the results of a semi-classical treatment of the problem based on discrete truncated Wigner approximation (DTWA)~\cite{Schachenmayer_dTWA2015,Kelly2021}, displaying qualitative agreement with NEQFT for large spins. However, DTWA predicts a slower relaxation for small $S$, possibly because as a semi-classical approach it ignores quantum fluctuations~\cite{POLKOVNIKOV_TWA2010}, which facilitate tunneling between local minima in the energy landscape of the SG.}

\begin{figure}[!t]
    \centering
    \includegraphics[width=.8\linewidth,trim={0 0.3cm 0 0.2cm},clip]{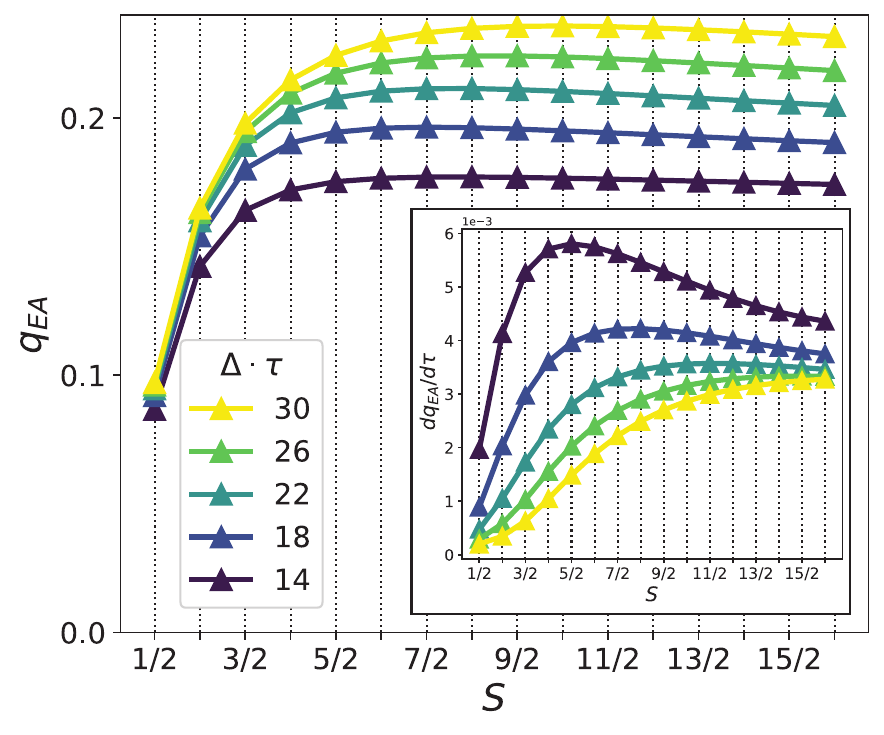}
    \caption{\blue{Effect of spin size on SG order. $q_\mathrm{EA}$ is suppressed by quantum fluctuations for small spins. It shows a sharp QC crossover at $S\approx2$ where it becomes weakly sensitive to $S$ and relaxes slowly to its final value. (Inset) Rate of change of SG order parameter. Compared to classical SG, quantum SG relaxes more quickly at initial times. At later times, dynamics become slow for all spin sizes, but they are faster for large spins as $q_\mathrm{EA}$ should still grow to reach steady state. The coupling is $g/\Delta=0.4$.}}
    \label{fig:EA_vs_Ns}
\end{figure}

The standard measure of SG order is the Edwards-Anderson \cite{Edwards_75,Buchhold_PRA13} order parameter given by the $t \to \infty$ limit of the symmetric correlation function
\begin{equation}\label{C}
     C(\tau,\tau-t)\equiv \frac{1}{N_s^2}\overline{\expval{\acomm{S_i^x(\tau)}{S_i^x(\tau-t)}}}.
\end{equation}
The system is a SG, if alongside with $m_x=0$, the quantity $q_\mathrm{EA}(\tau)\equiv\lim_{t \to \infty}C(\tau,\tau-t)$ is finite, which indicates a finite overlap between spin configurations at long time separations, a direct consequence of a frozen spin state.  We quench the system from a PM state with $\theta_0=\pi$ such that $m_x(0)=0$. As shown in Fig. \ref{fig:C_vs_g}, $C$ exhibits qualitatively different behaviors in PM and SG phases. In the former, the long time correlations are weak and the system reaches equilibrium quickly, as indicated by the lack of sensitivity to the waiting time $\tau$ after the quench. On the other hand, correlations decay very weakly with $t$ and show strong dependence on the waiting time for quenches into SG. Fig.~\ref{fig:C_vs_g} hints at the onset of aging, which is  a salient feature of glassy systems and an indirect signature of the breakdown of fluctuation-dissipation relations, or in other words of ergodicity~\cite{mezard1987spin,SG_RMP}.  In aging, the system strongly retains the memory of its past for infinitely long times after the quench. The inability to lose the memory of the initial state is a result of slow dynamics due to a rugged energy landscape generated by the disorder \cite{Cugliandolo_1994}.\\

\blue{\emph{Effect of spin size on SG dynamics} -- As previously discussed, the spin size $S=N_s/2$ is a crucial parameter which allows us to study the interplay of frustration and quantum fluctuations. We have already discussed the role of $S$ on the dynamics of global magnetization as a proxy of glassy behavior. In this section, we directly address quantum effects in the evolution of SG order parameter and subsequently, characterize quantum and classical SGs based on their dynamical signatures in a unified framework.
The correlation function $C$ is shown in Fig.~\ref{fig:C_vs_S} for different spin sizes, with a different dependence at small and long time separations ($t$). Short time ($t\lesssim \Delta^{-1}$) correlations are stronger for small spins, while they quickly flatten as $S$ grows. This behavior is easily explained by considering a generic local spin state $\ket{\psi}=\sum_{M}c_M\ket{M,S}$ with $S^x\ket{M,S}=M\ket{M,S}$. For $t\lesssim \Delta^{-1}$ we can approximate $C(t,t')\approx C(t,t)$ which yields $C(t,t)= \sum_M (M/S)^2 |c_M|^2 $. In SG phase, local magnetization is finite and $|c_M|^2$ has an asymmetrical distribution peaked around a finite value of $M$, leading to $C(t,t)\le 1$ where equality is approached only in the extreme case of a fully polarized state, or when $S$ is small. Particularly, for $S=1/2$ we always have $C(t,t)=1$, simply because the local Hilbert space is small. The slight deviation from the exact value for $S=1/2$ in Fig.~\ref{fig:C_vs_S} signals the validity of our approximation to capture quantum corrections to the dynamics. Long time correlations follow the opposite behavior; large spins are robust against quantum fluctuation and their correlation is weakly affected for $t\gtrsim \Delta^{-1}$. For $S\approx 1$, correlations decay sharply at $t\approx \Delta^{-1}$ which is the timescale of a single spin flip by the magnetic field (marked by the dashed line in Fig.~\ref{fig:C_vs_S}). Moreover, we see that long time correlations are drastically weaker for small $S$ compared to large $S$. Accordingly, the separation of short and long timescales for $S\approx 1$ and the lack of it for $S\gg 1$  motivate us to respectively label these as quantum and classical SG. This distinction is further supported by the dependence of SG order parameter $q_\mathrm{EA}$ on $S$ at different waiting times after the quench (Fig.~\ref{fig:EA_vs_Ns}). $q_\mathrm{EA}$ grows quickly with $S$ as the quantum noise is suppressed, and almost saturates for $S\gtrsim 2$, suggesting the system is already a classical SG (corresponding to $S\to \infty$). The growth of $q_\mathrm{EA}$ versus $S$ empirically admits an exponential fit of the form $q_\mathrm{EA} = q_c - r_q \, e^{- A S}$, where $q_c$ is the classical value and $r_q$ is the amplitude of quantum corrections which are exponentially suppressed in $S$. The quantities $q_c$, $r_q$ and $A$ depend on other parameters including the waiting time $\tau$. The exponential behavior holds up to larger values of $S$ as $\tau$ is increased (Fig.~\ref{fig:EA_vs_Ns}), because the order parameter growth is initially slower at larger $S$ and becomes faster later (inset of Fig.~\ref{fig:EA_vs_Ns}). The quick approach of $q_\mathrm{EA}$ to its steady state value for quantum SG in contrast to its slow growth for classical SG is another dynamical signature of the two regimes. A classical SG is glassier with a larger order parameter, stronger frustration and hence, slower dynamics.}

\begin{figure}[!t]
    \centering
    \includegraphics[width=.8\linewidth,trim={0 0.3cm 0 0.2cm},clip]{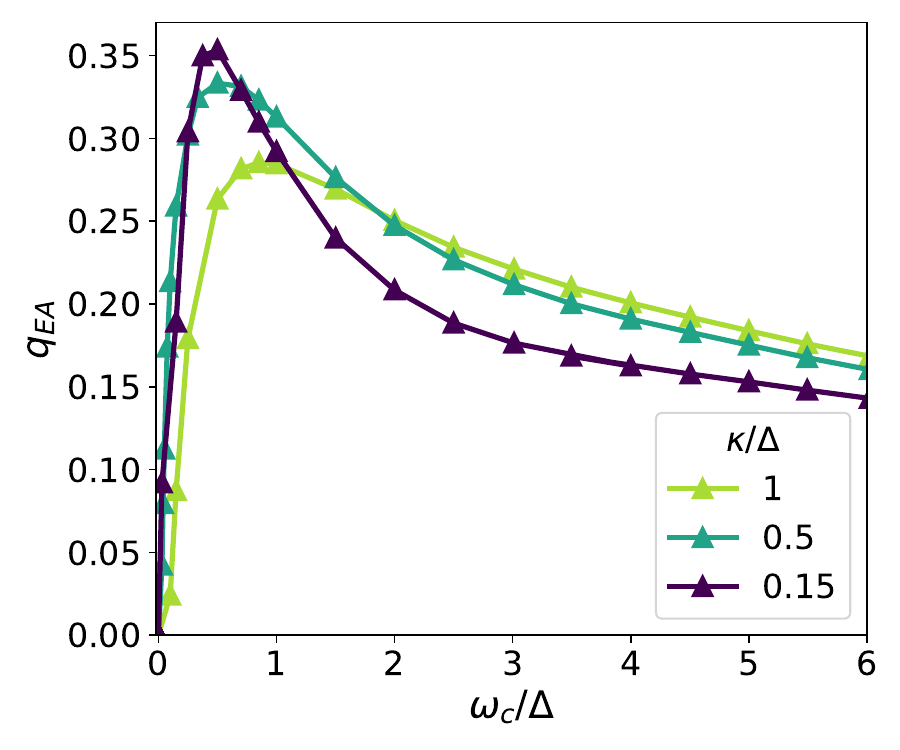}
    \caption{\blue{Resonantly enhanced SG order. In the limit $\omega_c\gg \Delta$ photons are coherently driven and SG order is independent of $\omega_c$. Near the resonance $\omega_c\approx\Delta$, SG order is amplified. For $\omega_c \lesssim \Delta$, photons are incoherently generated which leads to heating and weak SG. The resonance width scales with photon loss rate $\kappa$. The coupling is adjusted as $\omega_c$ is varied according to $g \approx 0.9 \sqrt{\Delta(\omega_c^2+\kappa^2)/\omega_c}$. The waiting time and the spin size respectively are $\Delta \tau = 12.0$ and $S=10$.}}
    \label{fig:resonance}
\end{figure}

\emph{Role of photons --} We now shortly discuss the effect of photons on the SG phase, as we change their frequency $\omega_c$. In the adiabatic limit ($\omega_c \gg \Delta$), the critical coupling depends on $\omega_c$ through $g_c \sim \sqrt{(\omega_c^2+\kappa^2)/\omega_c}$~\cite{Strack_PRL11}. Accordingly, we scale $g$ with $\omega_c$ similarly to $g_c$. In this way, when we compare physics at different values of $\omega_c$, the distance from the critical point remains fixed which enables us to isolate the effect of dynamically active photons (retardation effects). \blue{The dependence of SG order parameter on $\omega_c$ is shown in Fig.~\ref{fig:resonance}. For $\omega_c/\Delta \gg 1$, $q_\mathrm{EA}$ approaches a constant value, consistent with the adopted scaling of $g$. As $\omega_c$ is reduced, SG order gets enhanced until it reaches a maximum around $\omega_c \approx \Delta$ with a width that increases with photon loss $\mathcal{(\kappa)}$. Reducing $\omega_c$ further causes $q_\mathrm{EA}$ to vanish quickly over a small energy window. The peak in $q_\mathrm{EA}$ is the outcome of an intricate competition between two elements. First, the strength of photon-mediated interactions, which is resonantly enhanced as $\omega_c \to \Delta$. Second, the generation of incoherent photons by atoms which is limited by energy conservation, and is amplified when the energy scales of atomic and cavity excitations are close. Incoherent photons lead to heating, which subsequently weakens the SG order. The energy for heating is provided by the laser drive responsible for the emergence of the Dicke coupling in Eq.~\eqref{H}, after adiabatically eliminating virtual transitions to high energy intermediate states~\cite{mivehvar2021cavity,Marsh_PRX21,Marsh2023}. We see that the resonant enhancement dominates dissipation at $\omega_c\approx \Delta$, while as we reduce $\omega_c$ below the resonance, heating takes over and melts the SG. Energy conservation for incoherent photon generation has to be satisfied over a narrower energy window as cavity losses decrease, leading to a sharper resonance peak in Fig.~\ref{fig:resonance}. The location of the resonance peak does not exactly match $\omega_c=\Delta$ due to the renormalization and broadening of atomic and cavity energy levels by interactions. A similar resonance mechanism has been proposed recently ~\cite{eckhardt_resonance2024} to enhance superconductivity out of equilibrium, without controlling for heating. Our findings suggest the enhancement is robust with respect to heating effects.}\\

\emph{Method} -- To derive the results of this work, we used the Keldysh field theory~\cite{kamenev,Rammer_2007} extended to open quantum systems~\cite{Sieberer_2016}. Using a fermionic representation for spins~\cite{Schnirman_PRL03,Mao_PRL03} and employing a conserving approximation based on a quantum effective action (QEA)~\cite{Cornwall_74,Berges_02,berges2004introduction,calzetta2009nonequilibrium,Babadi_PRX15,Eberlein_PRB17,Haldar_PRR20,HH_PRB23,lang2023field,Bode_PRR24,gopalakrishna2023time} for the system, we dynamically monitored the onset of SG in the thermodynamic limit. We expanded QEA in powers of $1/N_s$, corresponding to successive atom-photon scattering processes~\cite{HosseinLong}. 
~\\

\blue{\emph{Perspectives} -- The distinction between quantum and classical SG dynamics addressed in this Letter holds beyond cavity QED platforms and encompass SG systems whose constituent degrees of freedom are the generators of the $SU(2)$ group~\cite{SG_RMP}. These include the canonical cases of infinite range quantum Ising and Heisenberg SG models, as well as the quantum $p$-spin model~\cite{SG_RMP,mezard1987spin}. On the other hand, the physics of frustrated systems with finite range interactions is more intricate and the interplay of quantum fluctuations, dimensionality and symmetries can change the behavior of QC crossover. 
For instance, in short range and low dimensional systems of small spins, quantum fluctuations may destabilize SG order towards a quantum spin liquid phase~\cite{balents_2010SL,savary_2016SL,zhou_2017SL}. This would represent a natural follow-up direction of our work. }

\blue{Furthermore, the QC crossover explored here should also manifest in the aging dynamics of quenches close to the FM or AFM critical points of quantum spin systems, without disorder~\cite{calabrese2005ageing,Gagel_prethermal14,Gagel_QAging15,Chiocchetta_prethermal17}.
 Finally, our findings concerning the resonant enhancement of SG order appear promising for systems in which a collective mode generates an effective interaction. The interaction can then be activated or amplified via an external drive to realize novel phases of matter without equilibrium counterparts such as those facilitated by driven excitons in doped Moir{\'e} systems~\cite{Yang_MoireExciton2023} or driven phonons in photo-enhanced superconductors~\cite{Knap_SC2016,Babadi_SC2017,eckhardt_resonance2024}. Our work suggests that the resonance condition should also be able to stabilize these emergent non-equilibrium phases~\cite{Kelly_resonance23,eckhardt_resonance2024}.}\\

\emph{Note Added} -- During the submission of this paper, we became aware of the recent experiment in Ref.~\cite{kroeze2023replica}  which reports, for the first time,   a spin glass   in the multi-mode cavity QED platform analysed here.\\

\begin{acknowledgements}
   \emph{Acknowledgements} -- We thank F. Balducci, J. Keeling and Riccardo J. Valencia-Tortora  for useful discussions, and D. Gribben for early contributions to this line of research. HH and JM acknowledge financial  support by the Deutsche Forschungsgemeinschaft (DFG, German Research Foundation): through  Project-ID 429529648, TRR 306 QuCoLiMa (“Quantum Cooperativity of Light and Matter”), and through TRR 288 - 422213477 (project B09).   This project has been supported by  the QuantERA II Programme that has received funding from the European Union’s Horizon 2020 research and innovation programme  under Grant Agreement No 101017733 ('QuSiED') and by the DFG (project number 499037529). DC acknowledges support from the European Union, under European Research Council grant agreement No 101002107 (NEWSPIN); the Government of Spain under the Severo Ochoa Grant CEX2019-000910-S [MCIN/AEI/10.13039/501100011033]); QuantERA II project QuSiED, co-funded by the European Union Horizon 2020 research and innovation programme (No 101017733) and the Government of Spain (European Union NextGenerationEU/PRTR PCI2022-132945 funded by MCIN/AEI/10.13039/501100011033); Generalitat de Catalunya (CERCA program and AGAUR Project No. 2021 SGR 01442); Fundaci\'{o} Cellex, and Fundaci\'{o} Mir-Puig. 
\end{acknowledgements}

\bibliography{refs}
\end{document}